
\input harvmac
\noblackbox
\def\ss{\sigma}
\def\gp{\gamma^+}

\def\tt{\tau}
\def\tth{\tilde \theta}
\def\th{\theta}
\def\PH{\Phi}
\def\xm{x^-}
\def\xp{x^+}
\def\xv{\vec x}

\def\Xv{\vec X}

\def\pp{p^+}
\def\pv{\vec p}

\def\dxp{\delta\xp}
\def\dxm{\delta\xm}
\def\dxv{\delta\xv}

\def\NP{{\it Nucl. Phys.\ }}

\def\PL{{\it Phys. Lett.\ }}
\def\PR{{\it Phys. Rev.\ }}

\def\IJMP{{\it Int. Jour. Mod. Phys.\ }}

\def\PRep{{\it Phys. Rep.\ }}
\baselineskip 12pt
\Title{\vbox{\baselineskip12pt\hbox{hep-th/9312107}\hbox{UCSBTH-93-39}
}}
{\vbox{\hbox{\centerline{Causal Properties of Free String Field Theory}}}}
\centerline{\it David A. Lowe\footnote*{lowe@tpau.physics.ucsb.edu}}
\centerline{Department of Physics}
\centerline{University of California}
\centerline{Santa Barbara, CA 93106-9530}
\smallskip
\vskip .5cm
\noindent
This paper examines the causal structure of the commutator
of two string fields, in free light-cone string field theory.
By treating the commutator as a distribution on
infinite dimensional loop space, it is shown that the
commutator vanishes when $\int d\ss (\delta X(\ss))^2 <0$.
Of more direct physical interest is the commutator of
finite mass fields, obtained by smearing the string fields
with appropriate wave functions. This is shown to vanish
at spacelike separations, reproducing the usual point
particle field theory result. The implications of this for
the information spreading mechanism proposed by Susskind
to solve the black hole information problem are discussed.
Finally, it is verified that the above conclusions
also hold for the superstring.

\Date{December, 1993}

\baselineskip 12pt
\newsec{Introduction}

Until recently, the notion of causality in string theory
has been relatively unexplored.
One might expect some kind of Planck scale violation
of causality due to the inherent non-local structure of
strings moving in space-time. This question was
addressed in
\nref\mart{E. Martinec, ``The Light Cone in String Theory,''
{\it Class. Quant. Grav.} {\bf 10} (1993) L187.}%
\mart\ where
the analog of the light-cone in string theory was
considered in the framework
\nref\mand{S. Mandelstam, \NP {\bf B64} (1973) 205; \NP {\bf B83}
(1974) 413; \PRep {\bf 13C} (1974) 259.}%
\nref\crem{E. Cremmer and J.-L. Gervais, \NP {\bf B76} (1974) 209;
\NP {\bf B90} (1975) 410.}%
\nref\lcstring{M. Kaku and K. Kikkawa, \PR {\bf D10} (1974) 1110;
\PR {\bf D10} (1974) 1823.}%
\nref\hop{J.F.L. Hopkinson, R.W. Tucker and P.A. Collins,
\PR {\bf D12} (1975) 1653.}%
of second-quantized bosonic light-cone string field theory
\refs{\mand{--}\hop}. In the second-quantized light-cone
formalism one has a consistent operator formulation,
so questions of local causality may be formulated in terms
of the commutator of fields. Note that the string field
in light-cone gauge only contains physical degrees of freedom,
so, in principle, it is an observable.
The calculations in \mart\ are valid when
one introduces either a lattice cutoff on the string
worldsheet or, equivalently, a cutoff on the number of string modes,
although such a cutoff is not explicitly introduced in \mart.
This is required
to avoid
divergences due to the infinite number of string states, which,
as shown in
\ref\green{E. Corrigan and D.B. Fairlie, \NP {\bf B91} (1975) 527;
M.B. Green, \NP {\bf B124} (1977) 461; \PL {\bf B266} (1991) 325;
\PL {\bf B302} (1993) 29; preprint QMW-91-02.}
are sufficient to cause the propagator of
point-like string states
to blow up at spacelike separations, even in the
superstring case. In this
paper, we show this kind of cutoff is unnecessary, as long
as one regards the commutator and the propagator as
distributions on loop space. Then it is clear that to make
well-defined statements one needs to smear these distributions
with smooth test functions. This procedure leads to a
simple proof of the vanishing of the commutator in the free theory
at the
string analog of spacelike separations. In
\ref\other{E. Martinec, "Strings and Causality,''
hepth/9311129, EFI-93-65.}
a weaker version of this condition was found, which appeared
to allow for interference between string fields at spacelike
separations.

To see what this causality condition means physically, it is
important to consider not just eigenstates
of the string localized in loop space,
but finite mass wave packets. As is now
well-known, these necessarily involve strings that extend over
all space
\ref\igor{M. Karliner, I. Klebanov and L. Susskind, ``Size
and Shape of Strings,'' \IJMP {\bf A3} (1988) 1981.}.
Nevertheless, we find these finite mass field operators commute
at spacelike separations as is usual in point-particle
quantum field theory. We argue that
this implies a na\"{\i}ve interpretation of the information
spreading mechanism recently suggested by Susskind
to explain the black hole information problem, does not appear to
occur, at least in flat-space, at the level of free string field
theory.

The calculations discussed above were performed in bosonic
string theory.
In order to formulate a well-defined causal boundary in the first place,
it was necessary to
adjust the mass of the tachyon by hand, with the
justification that in the full superstring field theory,
the physics would be essentially the same. As a check on this
we perform the calculation using second-quantized
light-cone superstring field theory,
where the string field lives
on superspace. It is verified that the boundary of causal propagation
is identical to that found in the bosonic string calculation.

\newsec{Causality and the Commutator in String Field Theory}

Here we will use the second quantized light-cone
string field theory developed in \refs{\mand{--}\hop}.
We introduce the light-cone coordinates
\eqn\lccord{
X^+ = (X^0 + X^{d-1})/ \sqrt{2}\qquad {\rm and} \qquad
X^- = (X^0 - X^{d-1})/ \sqrt{2}~.
}
Light-cone gauge corresponds to fixing $X^+(\ss) = \xp$. The
string field is then a function of the zero-modes
$\xp,\xm_0$ and the transverse
coordinates $\vec X(\ss)$. For simplicity we consider the
case of open strings -- the closed string case is very similar.
The transverse coordinates are expanded as
\eqn\xtrans{
\vec X(\ss) = \xv_0 + \sum_{l=1}^{\infty} 2 \xv_l \cos(l \ss)~.}
The string field is decomposed in terms of creation and annihilation
operators as
\eqn\pdecom{
\eqalign{
\Phi(\xp, \xm_0, \Xv(\sigma)) &=
\int_{-\infty}^{\infty} {{d\pv} \over {(2\pi)^{d-2}}}
\int_0^{\infty} { {d\pp} \over {2\pp}}
\sum_{ \{ \vec n_l \} } \biggl(
A_{\pp, \pv, \{ \vec n_l \} } e^{i(-\pp \xm_0 - p^- \xp +
\pv \cdot \xv_0)} \prod_{l=1}^{\infty}
f_{ \vec n_l}( \xv_l) \cr &+ h.c. \biggr)~.\cr}
}
Here the $A$ operators obey  canonical commutation relations
\eqn\commut{
[ A_{\pp, \pv, \{ \vec n_l \} }, A^{\dag}_{ {\pp}', {\pv}\,', \{
\vec n_l\,' \} } ] = 2 \pp \delta(\pp-{\pp}') (2\pi)^{d-2}
\delta^{d-2}(\pv -{\pv}\,') \delta_{ \{ \vec n_l \}, \{
\vec n_l\,' \} }~,
}
while the $f_{ \vec n_l}( \xv_l)$ are harmonic oscillator
wave functions given by
\eqn\harmon{
f_{ \vec n_l}( \xv_l) = \prod_{i=1}^{d-2} H_{ n_l^{(i)} }(x_l^{(i)} )
e^{ -l x_l^{(i)2} }~,
}
with $H_{ n_l^{(i)} }(x_l^{(i)} )$ a normalized Hermite polynomial. The
light-cone energy is defined by
\eqn\energ{
p^- = {{ \bigl( \pv^2 +m_0^2 + \sum_{l=0}^{\infty} l \vec n_l \bigr)}
\over {2 \pp}}~.
}
The string propagator is then
\eqn\sprop{
\eqalign{
&G(\xp_1, \xm_{1,0}, \Xv_1(\sigma); {\xp_2}, {\xm_{2,0}}, {\Xv_2}(\sigma) )
\equiv
\vev{0| \Phi(\xp_1, \xm_{1,0}, \Xv_1(\sigma)) \Phi({\xp_2}, {\xm_{2,0}},
{\Xv_2}(\sigma))
|0} = \cr &
\int_0^{\infty} { {d\pp} \over {2\pp}} e^{-i \pp(\xm_{1,0}-{\xm_{2,0}})}
 K^0_{\pp}(\xp_1, \xv_{1,0}; {\xp_2}, {\xv_{2,0}}) \prod_{l=1}^{\infty}
K^l_{\pp}(\xp_1, \xv_{1,l}; {\xp_2}, {\xv_{2,l}})~, \cr}
}
where we have defined
\eqn\hoscprop{
\eqalign{
K^0_{\pp}(\xp,\xv; {\xp}',{\xv}\,')
&=
\bigl( {{\pp} \over {2\pi \dxp}} \bigr)^{(d-2)/2} \exp i\biggl(
-{{m^2} \over {2 \pp}} \dxp + {{\pp} \over {2\dxp}} (\dxv)^2
 \biggl) ~, \cr
K^l_{\pp}(\xp_1, \xv_{1,l}; {\xp_2}, {\xv_{2,l}}) &=
\biggl( {{ 2 l} \over {\pi(1-\exp( -2i l \dxp /\pp)) } } \biggr)^{(d-2)/2}\cr
& \times \exp\biggl( { {-i l} \over { \sin( l \dxp/ \pp) }} \bigl(
(|\xv_{1,l}|^2 + |{\xv_{2,l}}|^2 ) \cos( l \dxp/\pp) -2 \xv_{1,l} \cdot
{\xv_{2,l}} \bigr) \biggr)~, \cr}
}
with $\dxp = \xp_1 -\xp_2$. The tachyon mass squared $m^2$ is to be
treated as a positive adjustable parameter for the purpose of
discussing  the causal properties of the commutator.
Of course, this is inconsistent with Lorentz invariance for
the bosonic string. The superstring will be treated in section
4, where $m^2=0$, to verify this procedure does indeed describe
the correct physics.

It is important to regard the string field as an operator-valued
distribution.
This means the string field should, in general, be smeared with
smooth test functions in order to obtain well-defined expressions.
Likewise the propagator must also
be regarded as a distribution. Now let us consider
the commutator of two smeared string fields. For the purpose
of seeing the causal properties of the commutator
it suffices to leave a finite number $N-1$ of the string modes
unsmeared. This leads to
\eqn\smcomm{
\eqalign{
\int \prod_{l=N}^{\infty}& d \xv_{1,l} d \xv_{2,l} g( \{ \xv_{1,l} \} )
h( \{ \xv_{2,l} \} ) [ \PH(X_1) , \PH(X_2) ] = \cr &
\int \prod_{l=N}^{\infty} d \xv_{1,l} d \xv_{2,l} g( \{ \xv_{1,l} \} )
h( \{ \xv_{2,l} \} ) \biggl( G(X_1; X_2) - G(X_2; X_1) \biggr)~, \cr}
}
where $g$ and $h$ are $C^{\infty}$, square integrable test functions.
We wish to
consider when we are allowed to rotate the $\pp$ contour
along either the positive or negative imaginary axis.
We begin by considering the first term in \smcomm, and
we assume $\dxp>0$. Rotating the $\pp$ integral along the
positive imaginary axis leads to a convergent integral
near $\pp=0$, however divergences may appear from the
$\pp\to \infty$ limit. In this limit, this term in \smcomm\
looks like
\eqn\inlim{
\eqalign{
\int^{\infty}& {{d\pp} \over {2 \pp}}
\int \prod_{l=N}^{\infty}  d \xv_{1,l} d \xv_{2,l}~g( \{ \xv_{1,l} \} )
h( \{ \xv_{2,l} \} )
e^{ \pp (\xm_{1,0} -\xm_{2,0})} \cr & \times
\biggl( { {\pp} \over {2\pi \dxp}} \biggr)^{(d-2)/2}
\exp\biggl(-{{\pp} \over {2\dxp}} (\delta \xv_0)^2 \biggr)
\prod_{j=1}^{\infty} \biggl( { {\pp} \over {\pi \dxp}} \biggr)^{(d-2)/2}
\exp\bigl(- {{\pp} \over {\dxp}} (\delta \xv_j)^2 \bigr) ~.\cr}
}
Note that in this limit the integrand is proportional to
an infinite product
of $\delta$-functions in $\delta \xv_j$. Integrating over the test
functions shows that this contribution to \smcomm\ is finite,
provided
\eqn\lccond{
2 \dxm_0 \dxp - \dxv_0^2 - 2\sum_{l=1}^{N-1} | \dxv_l|^2 < 0~.
}
Likewise, the contour in the second term of \smcomm\ may
be rotated along the negative imaginary axis provided
\lccond\ is satisfied. One then finds the two terms cancel
and the commutator vanishes when \lccond\ holds.
We may now note this argument did not depend on the value
of $N$, as long as it is finite, and did not depend on the
precise nature of the test functions, as long as they were
sufficiently smooth and square integrable.
Therefore, formally taking the
$N\to \infty$ limit,
it is consistent to regard the commutator of two
string fields as vanishing, in the sense of distributions, provided
\eqn\cond{
2 \dxm_0 \dxp - \dxv_0^2 - 2\sum_{l=1}^{\infty} | \dxv_l|^2 < 0~.
}

This condition is stronger than that found in \other\ which
stated that the commutator vanishes when
\eqn\wcond{
2 \dxm_0 \dxp - \dxv_0^2 - 2\sum_{l=1}^{\infty} | \dxv_l|^2 < -1~,
}
which means that operators may interfere at the string
analog of spacelike separations.
This condition \wcond\ appears inconsistent with the
equal $\xp$ canonical commutation relations of light-cone
string field theory which require
\eqn\cancom{
[ \PH(X), \PH(X') ] = \Delta(X(\ss)-X'(\ss))~,
}
where $\Delta$ is a functional $\delta$-function.
The source of the discrepancy lies in the treatment
of the infinite number of modes of the string. In this
paper we have smeared an infinite number of these modes
with smooth test functions. When we do this, no matter
what test functions we choose (as long as they satisfy the conditions
described above)
the interference at small
spacelike separations found in \other\ disappears.
We conclude, therefore, that \cond\ gives the correct
condition for the vanishing of the string field commutator,
when this object is properly interpreted as a distribution
on loop space.

\newsec{Physical Interpretation}

Let us see what the string light-cone condition
\cond\ tells us when we consider
finite mass wave packets of string fields.
It is well known
\igor\
that any finite mass superposition of string loops
necessarily involves strings of infinite length.
The
field operator that creates, for example, a gauge boson,
involves strings that occupy all of space. This is obtained
by
\eqn\photon{
A^k(\xp,\xm_0, \xv_0) =
\int \prod_{l=1}^{\infty} d\xv_l~f_{\vec n_l} (\xv_l) \PH(\xp,\xm_0, \xv_0,
\{ \xv_l\} )
}
with $x_0$ (the center of mass of the particle) fixed,
$\vec n_l=0$ for $l>1$ and $n_1^k=1$. Let us now
ask when two gauge boson field operators commute. Applying \cond\ to
each component of the $A$ field, one reproduces the usual
field theory result that two gauge boson fields always commute
at spacelike separations, i.e. when
\eqn\flight{
2 \dxp \dxm - | \dxv_0|^2 < 0~,
}
and otherwise they may interfere\footnote\dag{This result was
developed in conversation with S. Giddings.}. The same result holds
for any finite mass string states. Note that the functions
$f_{\vec n_l} (\xv_l)$ are not smooth in the limit $l \to \infty$,
as we required the test functions to be in the previous
section.
However, it may be verified by explicit calculation that relaxing
this constraint does not lead to any problems in computing
the commutator of two finite mass fields in the case at hand.

This would seem to indicate that the information contained in
a finite mass wave packet is not spread over all space
at the level of free string theory if
observations are performed at sufficiently small time resolutions,
as recently suggested by Susskind
\ref\suss{L. Susskind, Stanford preprint SU-ITP-93-18, July 1993;
SU-ITP-93-21, August 1993; Rutgers preprint RU-93-44, September 1993.},
but propagates just as in ordinary point-particle field theory.
To see this consider two observers ${\cal O}$ and ${\cal O}^{\prime}$
at spacelike separation.
Here we will follow closely an argument made in the
case of relativistic quantum field theory in
\ref\eber{P.H. Eberhard and R.R. Ross, {\it Found. Phys. Lett.} {\bf 2}
(1989) 127.}.
We make the hypothesis that probability distributions may be
computed using second-quantized string field theory. In addition,
we assume that a measurement performed at $x^{\mu}_0$ corresponds
to a measurement operator which is a function of finite mass
field operators
at $x^{\mu}_0$, and their derivatives. The corresponding projection
operators will then also be local in the finite mass fields, and
will commute at spacelike separations.
Near ${\cal O}$ is the center of mass
of some Planck mass string state (more precisely a localized wave packet
of such states). ${\cal O}$ then performs a local field strength measurement
which collapses the string state function. Now the question is,
if ${\cal O}^{\prime}$ performs a local
field strength measurement of sufficiently
small time resolution, does she see the change in the information
content of the system?

One interpretation of the results of \suss\ is that ${\cal O}^{\prime}$
should see the change in information, implying that superluminal
communication between ${\cal O}$ and ${\cal O}^{\prime}$ is possible.
However, because the projection operators
always commute at spacelike separations, we see that ${\cal O}^{\prime}$
is unable to measure the influence of ${\cal O}$ on the
state function. From this it seems the above interpretation of
the effect conjectured
in \suss\ does not occur, at least in flat Minkowski space, in free
string theory.

Intuitively, this results from the fact that the
wave function of a finite mass string
excitation (for example, a superposition of excitations up to
level $n$) has all the higher modes in their ground states.
It is this infinite number of higher modes which cause the
string wave function to spread over all space (in the
sense that $\vev{ ({\vec X}(\ss) -\xv_0)^2}$ diverges). However
it is clear these modes do not carry information since they are
in the same state for any finite mass excitation.

\newsec{Causality and the Commutator in Superstring Field Theory}

As a final check on the results of the previous sections,
we consider the generalization to the superstring in ten flat
space-time dimensions. In this theory, the worldsheet sweeps
out a surface in superspace, parametrized by $X(\ss,\tau)$ (the
space-time coordinate) and two Majorana-Weyl Grassmann coordinates
$\th(\ss,\tau)$ and $\tth(\ss,\tau)$.
We work in light-cone gauge where a consistent
${\rm SU(4)}\times {\rm U(1)}$ invariant superstring field
theory was developed in
\nref\gschwarz{M.B. Green and J.H. Schwarz, \NP {\bf B218} (1983) 43.}%
\nref\gsb{M.B. Green, J.H. Schwarz and L. Brink, \NP {\bf B219} (1983)
437.}%
\nref\restuccia{A. Restuccia and J.G. Taylor, \PRep {\bf 174} (1989) 283.}%
\ref\greens{M.B. Green and J.H. Schwarz, \NP {\bf B243} (1984) 475.}
\footnote*{See \refs{\gschwarz,\gsb} for earlier works on superstring
field theory, and \restuccia\ for a recent review.}.
This corresponds to the gauge choice
\eqn\lcgauge{
\eqalign{
X^+(\ss) &= \xp, \cr
\gp \th &= \gp \tilde \th = 0. \cr }
}
Here plus and minus refer to the longitudinal light-cone coordinates
defined by
\eqn\vlc{
V^{\pm} = (V^0 \pm V^9)/\sqrt{2}~.
}
In addition to the zero modes, the string wave function depends
on the transverse ${\rm SO(8)}$ vector $X^I(\ss,\tt)$ and the pair
of ${\rm SO(8)}$ spinor coordinates $\tt^a(\ss,\tt)$ and
$\tth^a(\ss,\tt)$. For chiral theories (type I and IIB) the
$\th$'s are in the same spinor representation ${\bf 8}_S$,
while in the non-chiral theory (type IIA) we choose
$\th^a$ to be in ${\bf 8}_S$ and $\tth^a$ to be in ${\bf 8}_C$.
In the ${\rm SU(4)}\times {\rm U(1)}$ formalism, the ${\rm SO(8)}$
spinor $\th^a$ is decomposed into the ${\rm SU(4)}$ spinors
$\th^{\bar A}$ and $\lambda^A$, where $A=1,2,3,4$, and likewise
for $\tth^a$.

For simplicity, let us restrict attention to the type I open string.
Yang-Mills charges may be attached to the ends of the string
for the groups SO(N) and USp(2N) as in
\ref\chan{J.E. Paton and H.M. Chan, \NP {\bf B10} (1969) 516.}.
The SO(32) theory is the
only open-string theory free of anomalies at the quantum level.
The string field belongs to the adjoint representation
of the gauge group.
The string field satisfies a non-orientability constraint
\eqn\noconst{
\PH^{ab}[ X(\ss), \th(\ss), \tth(\ss) ] = - \PH^{ba}[
X(2\pi |\pp| - \ss), \tth(2\pi |\pp| - \ss), \th(2\pi |\pp| - \ss) ]~.
}
In the following we set $Z(\ss) = ( X(\ss), \th(\ss), \tth(\ss))$.
The string field also satisfies a reality condition which
requires the $\PH$ field to be CTP self-conjugate. This
means
\eqn\ctp{
\hat \PH^{ab}(X,\th,\tth) = \PH^{ab*}(X,\th/4\pp, \tth/4\pp)
}
which, for example, restricts the field content of the
massless open-string states to the super Yang-Mills multiplet.
Here the symbol $\hat {\ }$ denotes the functional
Fourier transform with respect to the Grassmann
variables
\eqn\fourhat{
\hat \PH (X,\lambda,\tilde \lambda) = \int D^4 \th D^4 \tth
e^{I(\lambda,\th)} \PH(X,\th,\tth)~,
}
where
we have defined
\eqn\iexpo{
I(\lambda, \th) = \int_0^{2\pi \pp} d\ss
\biggl( \lambda^A(\ss) \th^{\bar A}(\ss) + \tilde \lambda^A(\ss)
\tth^{\bar A}(\ss) \biggr)~.
}
{}From this point  on we will suppress the group theory indices.
The equal $\xp$ commutation relations for the string fields satisfying
\noconst\ are
\eqn\cancom{
\biggl[ \PH[Z_1(\ss)] , \PH[Z_2(\ss) ] \biggr]=
2 \pp_2 \delta(\pp_1+\pp_2)
 \Delta^{16}[ Z_1(\ss) -Z_2(\ss)] ~.
}
Alternatively, one could relax \noconst\ for the fields in the
commutator, and subtract a piece proportional to
\eqn\expl{
\eqalign{
\Delta^{16}[ Z_1(\ss)-Z_2(2\pi |\pp_2|-\ss) ] &=
\Delta^8[ X_1^I(\ss) - X_2^I(2\pi |\pp_2|-\ss)] \cr & \times
\Delta^4[\th_1^{\bar A}(\ss) - \tth_2^{\bar A}(2\pi |\pp_2|-\ss)]
\Delta^4[\tth_1^{\bar A}(\ss) - \th_2^{\bar A}(2\pi |\pp_2|-\ss)] \cr}
}
from the right-hand side of \cancom, as done in \gschwarz.
The free action for the string field in light-cone gauge is
\eqn\action{
\eqalign{
S&=\int d\xp \int d^{16}Z \int_0^{\infty} {{ d\pp }\over {2\pp}}
-i  {\rm tr}( {\dot \PH}(Z,-\pp,\xp) \PH(Z,\pp,\xp) ) \cr
&-
{\rm tr}( \PH(Z,-\pp,\xp)  \int_0^{2\pi \pp} d\ss (
-\pi ({{\delta} \over {\delta X}})^2 + {1\over {\pi}} X^{\prime 2}
-2 i ( \th^{\prime} {{\delta} \over {\delta \th}} - \tth^{\prime}
{{\delta} \over {\delta \tth}} )  ) \PH(Z,\pp,\xp) ) ~.\cr}
}

We impose the open string boundary conditions
\eqn\bound{
\eqalign{
X'(0)&=X'(2\pi |\pp|)=0 ~,\cr
\th^{\bar A}(0) &= \tth^{\bar A}(0)~, \cr
\th^{\bar A}(2\pi |\pp|) &= \tth^{\bar A}(2 \pi |\pp|) \cr}
}
and use the following modal decompositions
\eqn\decom{
\eqalign{
X(\ss) &= x_0 + \sum_{n=1}^{\infty} 2 x_n \cos(n \ss/2 \pp) ~, \cr
\th^{\bar A}(\ss) &= {1\over {2\sqrt{2} \pp}} \sum_{n=-\infty}^{\infty}
R_n^{\bar A} e^{i n \ss/2 \pp} \cr
\tth^{\bar A}(\ss) &= {1 \over {2 \sqrt{2}  \pp}} \sum_{n=-\infty}^{\infty}
R_n^{\bar A} e^{-i n \ss/2 \pp} ~,\cr }
}
where the $R_n^{\bar A}$ should be regarded as Grassmann numbers.
In terms of these modes, the Schr\"odinger equation for the
string field is
\eqn\schro{
i { {\del \PH} \over {\del \xp}} =
\biggl( \sum_{n=1}^{\infty} \biggl[ {{n^2 x_n^2} \over { \pp}} -
{1\over {4\pp}} \bigl( {{\delta} \over {\delta x_n}}\bigr)^2 \biggr] -
{1\over {2\pp}}\bigl({{\delta} \over {\delta x_0}}\bigr)^2 -
{1\over {\pp}} \sum_{m=-\infty}^{\infty} m R_{m}^{\bar A} {{\delta} \over
{\delta R_{m}^{\bar A} }}
\biggr) \PH~.
}
The propagator obtained from this equation is
\eqn\sprop{
\eqalign{
G(Z_1,Z_2) &= \int_0^{\infty} {{d\pp} \over {2\pp}}
K^0_{\pp}(\dxp, \dxm, \dxv) \prod_{a=1}^{\infty} K^a_{\pp}(\dxp,
\xv_{1,a}, \xv_{2,a}) \cr &\times \prod_{b=-\infty}^{\infty} L^b_{\pp}(\dxp,
R_{1,b}, R_{2,b} ) ~.\cr}
}
Here $K^0_{\pp}$ is a free, mass $m$, particle propagator
\eqn\partpro{
K^0_{\pp}(\xp,\xv; {\xp}',{\xv}\,')
=
\bigl( {{\pp} \over {2\pi \dxp}} \bigr)^{(d-2)/2} \exp i\biggl(
-{{m^2} \over {2 \pp}} \dxp + {{\pp} \over {2\dxp}} (\dxv)^2
 \biggl) ~,
}
the $K^a_{\pp}$ are bosonic harmonic oscillator
propagators,
\eqn\bgreen{
\eqalign{
K^a_{\pp}(\dxp,\xv_{1,a}, \xv_{2,a}) &=
\biggl( { {2 a} \over {\pi (1-\exp(-2i a \dxp/ \pp) )}}
\biggr)^{(d-2)/2} \cr & \times
\exp\biggl( { {-ia} \over { \sin(a \dxp/ \pp) }} \bigl(
(|\xv_{1,a}|^2 + |\xv_{2,a}|^2 )\cos( a \dxp/ \pp) - 2 \xv_{1,a} \cdot
\xv_{2,a} \bigr) \biggr) ~,\cr }
}
while the $L^b_{\pp}$ are a representation
of the Fermi oscillator propagators
\eqn\rgreen{
\eqalign{
L^b_{\pp}(\dxp,R_{1,b}, R_{2,b} ) &=
R_{2,b} \exp( -i b \dxp/ \pp) - R_{1,b}\quad {\rm for } ~b\ge 0~,\cr
L^b_{\pp}(\dxp,R_{1,b}, R_{2,b} ) &=
R_{2,b}  - \exp(-i |b| \dxp/ \pp) R_{1,b}\quad {\rm for } ~b< 0~.\cr}
}
In the superstring $m^2=0$, however to separate poles in the
propagator we will take $m^2=0^+$.

Now we are ready to consider the commutator of the string
field at different times
\eqn\sfcomm{
\eqalign{
\biggl[ & \PH[Z_1(\ss),\xp_1], \PH[Z_2(\ss),\xp_2 ] \biggr]= \cr &
\int_0^{\infty} {{d\pp} \over {2\pp}} \biggl(
K^0_{\pp}(\dxp, \dxm, \dxv) \prod_{a=1}^{\infty} K^a_{\pp}(\dxp,
\xv_{1,a}, \xv_{2,a}) \prod_{b=-\infty}^{\infty} L^b_{\pp}(\dxp,
R_{1,b}, R_{2,b} ) \cr &-
K^0_{\pp}(-\dxp, -\dxm, -\dxv) \prod_{a=1}^{\infty} K^a_{\pp}(-\dxp,
-\xv_{1,a}, -\xv_{2,a}) \prod_{b=-\infty}^{\infty} L^b_{\pp}(-\dxp,
-R_{1,b}, -R_{2,b} ) \biggr) ~.\cr}
}

One may make the same kind of argument used
above to determine when the commutator of two string
fields vanishes. Again it is essential to regard
the commutator as a distribution-- this time on super-loop space
$Z(\ss)$. When smeared by smooth functions
the Grassmann factors make no contribution to the convergence in
the $\pp \to \infty$ limit--
the convergence is dominated only by the factors depending on the
bosonic coordinates $\xv_a$ as before. One therefore reaches
the same conclusion: the commutator of two string fields
vanishes if
\eqn\slight{
2 \dxp \dxm - | \dxv_0|^2 - 2 \sum_{a=1}^{\infty} | \dxv_a|^2 < 0~.
}
Note the natural space-time Lorentz covariant form of this
condition is
\eqn\cslight{
\int d\ss ( \delta X(\ss) )^2 < 0~.
}

\newsec{Conclusions}

In this paper we have carefully reconsidered the
derivation of the light-cone condition on string loop
space, first considered in \mart. By properly treating
the string commutator as a distribution on loop space
we have shown that the light-cone condition originally found
in \mart\ is correct, which is a stronger condition than that found
in a later calculation \other.
These calculations were performed
in free string field theory.
An interesting open question is whether interactions,
due to their non-local nature in string theory,
will change this light-cone condition in the full
\nref\lsu{D.A. Lowe, L. Susskind, J. Uglum, in preparation.}%
theory.\footnote\dag{Note added: In fact, this now appears to be the case
\lsu.}

Applying this causal condition to finite mass field
operators, it was shown that the usual point-particle
quantum field theory light-cone condition  is recovered.
It was argued that
the information spreading mechanism proposed in \suss\
to solve the black hole information paradox, does not occur
at the level of free string field theory.

As a final check on the consistency of the above calculations,
the causal condition was formulated in the framework
of light-cone superstring field theory. The bosonic
string contains a tachyon, whose mass must be set to zero
by hand in the analogous bosonic calculations. Therefore,
any completely consistent discussion of causality should
be made in the context of the superstring. It was shown, nevertheless,
the same causal condition is found.

\vskip 1in
{\bf Acknowledgments:} The author wishes to thank S. Chaudhuri,
S. Giddings, J.  Polchinski,
T. Samols and A. Strominger
for helpful discussions, and M. Green for pointing out
reference \green. This work is supported in part by
NSF grant PHY91-16964.

\vfill \eject
\listrefs
\end